\documentclass[twocolumn]{aastex701}

\usepackage{amsmath}
\usepackage{bm}
\usepackage{multirow}
\usepackage{mathrsfs}
\def\vec#1{\ensuremath{\bm{#1}}}
 
\begin{document}

\title{
Alfv\'enic Motions in a Stratified Open Flux Tube: Transition from Propagating to Locally Standing Motions and Implications for the Kelvin-Helmholtz Instability}

\author[orcid=0000-0003-4956-6040,sname='Guo']{Mingzhe Guo}
\affiliation{Institute of Frontier and Interdisciplinary Science, Shandong University, Qingdao 266237, China}
\affiliation{Shandong Key Laboratory of Space Environment and Exploration Technology, Institute of Space Sciences, Shandong University, Shandong, China}
\email[show]{m.guo@email.sdu.edu.cn}  
\correspondingauthor{Mingzhe Guo}

\author[orcid=0000-0003-4790-6718, sname='Li']{Bo Li} 
\affiliation{Shandong Key Laboratory of Space Environment and Exploration Technology, Institute of Space Sciences, Shandong University, Shandong, China}
\email{}

\author[orcid=0000-0002-6641-8034, sname='Gao']{Yuhang Gao} 
\affiliation{School of Earth and Space Sciences, Peking University, Beijing 100871, China}
\affiliation{Centre for mathematical Plasma Astrophysics (CmPA), Department of Mathematics, KU Leuven, Celestijnenlaan 200B, 3001 Leuven, Belgium}
\email{}

\author[orcid=0000-0001-6449-8838,sname='Chen']{Yao Chen}
\affiliation{Institute of Frontier and Interdisciplinary Science, Shandong University, Qingdao 266237, China}
\affiliation{Shandong Key Laboratory of Space Environment and Exploration Technology, Institute of Space Sciences, Shandong University, Shandong, China}
\email{}

\begin{abstract}
{
Standing transverse waves in closed coronal structures have been widely studied as a possible route to energy dissipation, 
with resonant absorption transferring kink wave energy to localized Alfv\'enic motions and the Kelvin-Helmholtz instability (KHI) accelerating the formation of small dissipative scales.}
However, it remains unclear whether the same mechanism applies to the open corona, given the long-standing consensus that the KHI tends to be prohibited for propagating Alfv\'enic waves. 
Within the framework of magnetohydrodynamics (MHD), 
we perform three-dimensional MHD simulations of boundary-driven kink waves in a gravitationally stratified open flux tube extending from the chromosphere into the corona. 
We find that propagating waves in open 
    magnetic structures can also drive the system toward a turbulent state,
    with KH vortices clearly identifiable across the flux tube. 
This occurs because resonant absorption transfers energy from the propagating kink waves to azimuthal Alfv\'enic motions near the tube boundary, and wave reflection off the gradient of the Alfv\'en speed 
subsequently enables these boundary motions to acquire a locally standing character. 
Our results provide a possible answer to the long-standing question 
of whether and how propagating waves in open magnetic structures can generate nonlinear turbulent fine structures despite their globally propagating nature.

\end{abstract}

\section{Introduction} 
\label{sec:intro}

Propagating Alfv\'enic waves are ubiquitously observed in open magnetic structures of the solar atmosphere. 
In open magnetic-field regions, observational signatures of such waves include persistent Doppler velocity fluctuations in the off-limb corona \citep[e.g.,][]{2007Sci...317.1192T}, direct transverse displacements in coronal plumes \citep[e.g.,][]{2014ApJ...790L...2T}, and propagating kink waves connecting the chromosphere and corona in plumelet structures \citep{2026ApJ..1001..173Q}. 
These waves are widely considered to play an important role in the dynamics and energetics of magnetised plasma structures,
and their energy transport and dissipation have been discussed in the context of coronal heating \citep[see][for recent reviews]{2020ARA&A..58..441N,2020SSRv..216..140V,2021SSRv..217...73N}. 
An essential issue is therefore how the energy carried by large-scale propagating waves can be transferred to smaller spatial scales, 
where dissipation and heating can become more efficient.

For propagating waves, 
this transfer to small scales is not straightforward. 
In a purely propagating transverse wave, 
the boundary velocity shear is transported upward together with the wave pattern and therefore does not remain at a fixed height. 
The magnetic perturbation also propagates with the velocity perturbation, carrying the boundary deformation along the tube. 
This is different from standing transverse waves, for which resonant absorption can convert the energy of a collective kink mode into localized Alfv\'enic motions in an inhomogeneous boundary layer \citep[see][for a review]{2011SSRv..158..289G}. 
These localized motions are subject to phase mixing \citep[e.g.,][]{1983A&A...117..220H}, 
which enhances transverse gradients and boundary velocity shear. 
The resulting shear can trigger the Kelvin-Helmholtz instability (KHI), 
producing vortical fine structures and facilitating the cascade of wave energy to smaller scales \citep[e.g.,][]{2008ApJ...687L.115T,2014ApJ...787L..22A,2017A&A...604A.130K,2019ApJ...870...55G,2019ApJ...883...20G,2021ApJ...908..233S,2023ApJ...949L...1G}. 
For propagating kink waves, however, the associated shear and boundary deformation are generally expected to be less persistent, 
which helps explain why propagating kink waves are generally expected to produce only limited phase-mixing-induced fine structures. 
Nevertheless, open magnetic structures in the solar atmosphere are both transversely structured and longitudinally stratified, 
and these inhomogeneities may modify the nonlinear evolution of propagating waves. 
For example, \citet{2024A&A...689A.195G} studied propagating kink waves in a stratified open coronal flux tube and found the initiation of KHI-like boundary structures, 
although these structures remained limited and did not develop into a fully turbulent state.

Therefore, whether propagating waves in open magnetic structures can generate well-developed KH vortices, and under what physical conditions this may occur, remains an open question. In this paper, we model an open magnetic flux tube extending from the chromosphere to the corona. We aim to investigate how propagating kink waves interact with transverse and longitudinal inhomogeneities, and whether such interactions can drive the system toward a turbulent state.
The paper is organized as follows. 
Section~\ref{sec:model} describes the model, including the equilibrium 
and numerical setup. 
In Section~\ref{sec:results}, 
we present the numerical results. 
Section~\ref{sec:discussion} discusses their implications.
Section~\ref{sec:sum} summarizes our findings.

\section{Model Description} \label{sec:model}

We model a straight magnetic flux tube embedded in a gravitationally stratified solar atmosphere extending from the chromosphere to the corona. 
The basic parameters are similar to those used by \citet{2023ApJ...949L...1G} and \citet{2026ApJ..1001..173Q}. 
In a Cartesian coordinate system $(x,y,z)$, 
the initial magnetic field is uniform and vertical in the entire domain,
$\vec{B}_0=B_0\hat{z},$
with $B_0=20~{\rm G}$. 
The gravitational acceleration is prescribed as $\vec{g}(z) = -g_{\odot}R^2_{\odot}/(z+R_{\odot})^2\hat{z}$,
where $g_\odot$ and $R_\odot$ are the solar surface gravity and solar radius, respectively.
The transverse and vertical temperature profiles are prescribed following 
\citet{2023ApJ...949L...1G}, 
except that the internal coronal temperature is set to $T_{\rm i}=1.8~{\rm MK}$ in the present model. 
The initial density and pressure are obtained from hydrostatic balance together with the ideal gas law,
assuming a fully ionized hydrogen plasma.
The flux tube is aligned with the $z$-axis, 
has a radius $R=1~{\rm Mm}$, and extends from $z=0$ to $z=100~{\rm Mm}$. 
{ Since the internal temperature $T_{\rm i}$ is lower than the external temperature, $T_{\rm e}=3.6$ MK, }
the flux tube is density-enhanced with respect to the ambient corona. 
In addition, 
because the chromospheric temperature is taken to be uniform, 
we impose an initial pressure contrast of $p_{\rm i}/p_{\rm e}=3$ in the chromosphere, 
which also corresponds to a chromospheric density contrast of 3.
This setup leads to a relatively larger density contrast in the coronal part than that in 
\citet{2023ApJ...949L...1G}. 
The transition region is artificially broadened through a modified field-aligned 
thermal conductivity,
{ following the scheme developed by \cite{2001ApJ...546..542L}, \cite{2009ApJ...690..902L}, \cite{2013ApJ...773...94M}}, as described in detail by \citet{2023ApJ...949L...1G} and \citet{2023ApJ...955...73G}.

{
The present configuration is intended to represent an idealized, 
density-enhanced open magnetic flux tube rooted in the chromosphere, 
rather than a realistic polar plume or an individual spicule. 
Its diameter, $2R=2~{\rm Mm}$, is smaller than the characteristic widths reported for observed plumelets and may therefore be interpreted as an elementary strand within a plume-like open magnetic structure \citep[e.g.,][]{2021ApJ...907....1U}. 
The adopted internal coronal temperature
is somewhat higher than that commonly inferred for classical polar plumes \citep[e.g.,][]{2011A&ARv..19...35W,2015LRSP...12....7P}. 
The prescribed pressure contrast, $p_{\rm i}/p_{\rm e}=3$, therefore also corresponds to an initial density contrast of $\rho_{\rm i}/\rho_{\rm e}=3$ in the chromosphpere.
This representative value is adopted to construct a density-enhanced chromospheric root and provides a sufficiently clear transverse Alfv\'en speed contrast for the propagation and resonant coupling of kink waves. 
It is not intended as a quantitative representation of a particular observed plume or spicule. In addition,
the pressure and density distributions are subsequently modified during the relaxation, and the resulting density contrasts are listed in Table~\ref{tab1}.
These idealizations allow us to isolate the effects of 
gravitational stratification and transverse density structuring on the propagation and nonlinear evolution of kink waves.
}

We consider an enhanced-viscosity dissipative buffer to absorb velocities in the upper part of the domain to avoid potential reflection.
The explicit artificial viscosity in the entire computational domain is prescribed as
\begin{equation}
\nu(z)=\nu_{\rm buf} h(z),
\label{eq_nu}
\end{equation}
where $\nu_{\rm buf}$ is the enhanced value in the buffer.
It is given by $\nu_{\rm buf}=l_{\rm c}v_{\rm c}/{\rm Re}_{\rm buf}$, 
with $l_{\rm c}$ being the characteristic length of 1000 km and $v_{\rm c}$ being the typical velocity of 1000 km/s.
The Reynolds number Re is of order of $10$,
which ensures a much larger viscosity than the numerical one. 
The transition function $h(z)$ is chosen to ensure the smooth buffer,
 it reads
\begin{equation}
h(z)=1-\frac{1}{2}\left\{1-\tanh\left[ 0.1\left(z-z_{\rm buf}\right)\right]\right\},
\label{eq_buffer}
\end{equation}
where $z_{\rm buf}=50~{\rm Mm}$. 
Note that in practical runs, 
the viscosity changes from ${\rm Re}_{\rm buf}$ in the dissipative buffer to numerical 
Reynolds number in the main physical domain,
which is much larger than the ${\rm Re}_{\rm buf}$.
This layer damps upward-propagating waves before they reach the closed upper boundary, 
thereby reducing artificial reflection without directly modifying the velocity field.
The effectiveness of this buffer will be discussed in the following text.

The boundary conditions are specified as follows.
At the bottom boundary $z=0$, density and pressure are extrapolated from the hydrostatic equilibrium, 
and the magnetic field is extrapolated following a zero-normal-gradient condition \citep[see also,][]{2023ApJ...949L...1G,2019A&A...623A..53K}.
The vertical velocity $v_z$ is fixed to zero, while transverse perturbations are imposed to excite waves in the tube. 
Following the prescription in \cite{2019ApJ...870...55G},
a kink-like driver is employed,

\begin{equation}
\vec{v}(x,y,z;t)=
\left[
\vec{v}_{\rm e}
+\dfrac{1}{2}\left(\vec{v}_{\rm i}-\vec{v}_{\rm e}\right) f(r)
\right]
\sin\left(\dfrac{2\pi t}{P}\right),
\label{eq:v_driver}
\end{equation}
where 
\begin{equation}
\vec{v}_{\rm i}=v_0\hat{x},
\vec{v}_{\rm e}=v_0\left[\left(\frac{{x'}^2-y^2}{r^4}\right)\hat{x}
               +\left(\frac{2x'y}{r^4}\right)\hat{y}\right],
\label{eq_v_ie}
\end{equation}
and 
\begin{equation}
x'=x+\left(\frac{v_0P}{2\pi}\right)\cos\left(\frac{2\pi}{P}t\right),
r = \sqrt{{x'}^2+y^2}.
\label{eq_xper}
\end{equation}
where $v_0=5$ km/s and $P=100$ s are the driver amplitude and period, 
and $f(r)=1-\tanh\left[b(r/R-1)\right]$ specifies the transverse profile of the driver. 
Such amplitude and period are commonly confirmed in observations in the chromosphere \citep[e.g.,][]{2013ApJ...768...17M,2026ApJ..1001..173Q}.
All lateral boundaries are treated as outflow boundary conditions. 
{ At the upper boundary, all velocity components are fixed to zero, while the density, gas pressure, and magnetic field components are fixed to their relaxed values. 
The boundary is therefore formally closed and reflecting.
Nevertheless,
the dissipative layer can help suppress possible reflection into the physical domain. 
}

We solve the three-dimensional MHD equations with anisotropic thermal conduction and explicit viscosity using the PLUTO code \citep{2007ApJS..170..228M}. 
No explicit resistivity is included. 
Spatial reconstruction is performed with a piecewise linear method, numerical fluxes are computed using the Roe Riemann solver, and time integration is carried out with a second-order characteristic tracing scheme. The hyperbolic divergence-cleaning method is used to control $\nabla\cdot\mathbf{B}=0$. 
The computational domain is
$
[-6,6]~{\rm Mm}\times[-3,3]~{\rm Mm}\times[0,100]~{\rm Mm},
$
and is resolved by
$
N_x\times N_y\times N_z=256\times256\times512
$
uniform cells, corresponding to a transverse resolution of about $46.9~{\rm km}$. 

Note that relaxation is necessary, 
as the system is not magnetostatic balance.
After a relaxation period of about 8.71 ks,
the system reaches equilibrium with a maximum velocity below 1 km/s.
{
The density contrast varies after relaxation. 
The internal and external densities of the relaxed atmosphere at representative heights are listed in Table~\ref{tab1}.
The corresponding density ratios are broadly consistent with the 
plume-to-interplume density contrasts inferred in the lower corona \citep[e.g.,][]{2011A&ARv..19...35W}, 
although the present model is not intended to quantitatively reproduce a particular observation.
}

\begin{table}
\caption{{ Internal and external number densities in the relaxed atmosphere. Index i (e) denotes internal (external) values.}}
\begin{tabular}{cccc}
\hline\hline
$z$ & $n_{\rm i}$ & $n_{\rm e}$ & $n_{\rm i}/n_{\rm e}$ \\
    & (${\rm cm}^{-3}$) & (${\rm cm}^{-3}$) & \\
\hline
$10~{\rm Mm}$ & $1.6\times10^{10}$ & $6\times10^{9}$ & 2.67 \\
$20~{\rm Mm}$ & $1.1\times10^{10}$ & $3.8\times10^{9}$ & 2.89 \\
$40~{\rm Mm}$ & $7.1\times10^{9}$ & $2.5\times10^{9}$ & 2.84 \\
\hline
\end{tabular}
\label{tab1}
\end{table}

\section{Results} \label{sec:results}

\begin{figure*}[ht!]
	\centering
	\gridline{\fig{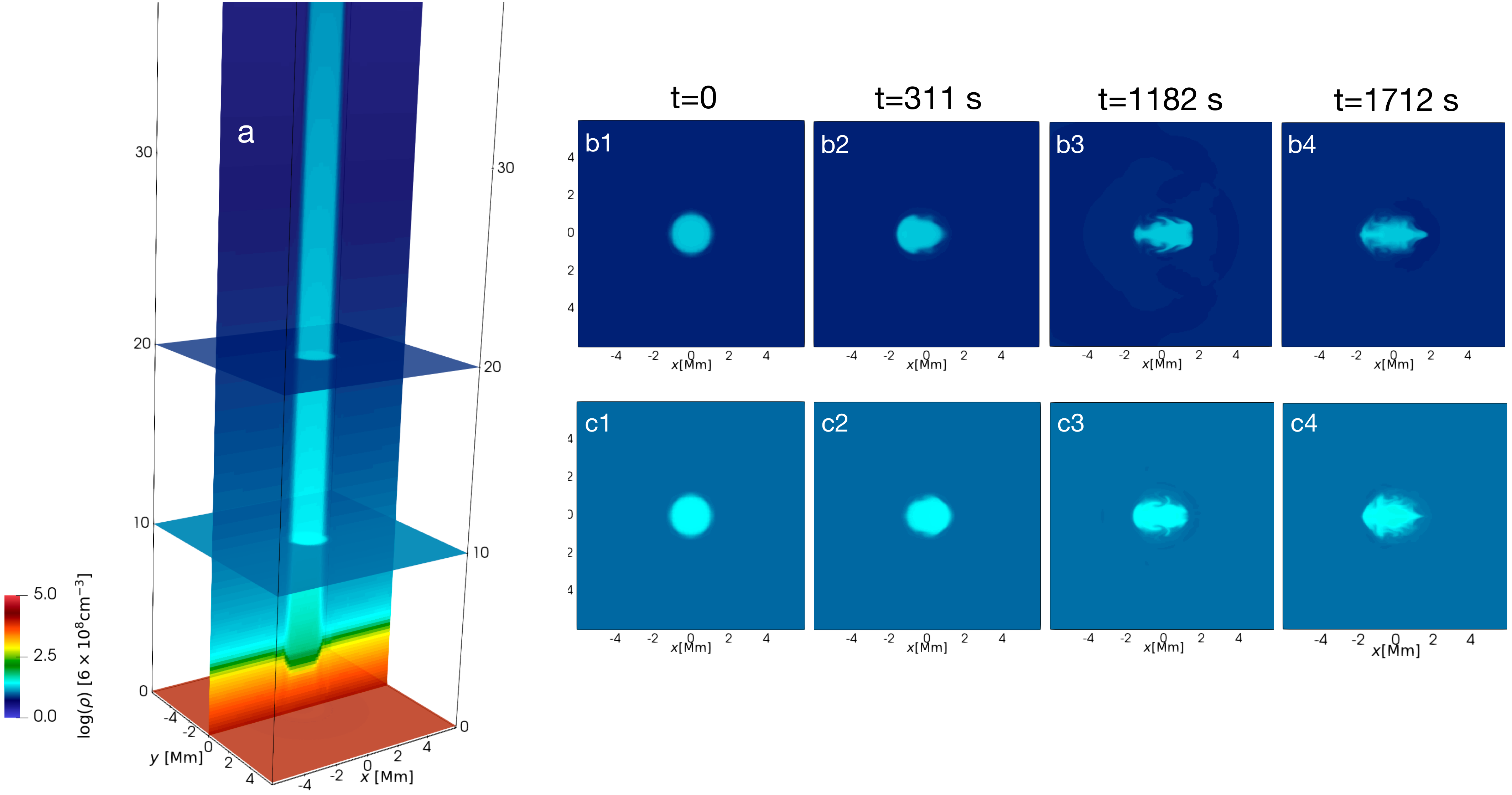}{1.0\textwidth}{}}
\caption{
{
(a) Selected slices of the relaxed state of the 3D magnetic flux tube embedded in a stratified atmosphere. 
Panels (b) and (c) show the temporal evolution of the tube cross section at $z=10~{\rm Mm}$ and $z=20~{\rm Mm}$, respectively, at the indicated times. 
An animated version of this figure is available in the HTML version of the article. 
The animation shows the continuous evolution of the tube cross sections from $t=0$ to 1742 s, 
with the simulation time indicated in each frame. 
Whereas the static panels show selected snapshots, 
the animation illustrates the progressive deformation of the tube boundary and the development of small scale KH vortices.
}
\label{fig:crs}}
\end{figure*}

We first examine the dynamics of the tube cross section. 
Figure~\ref{fig:crs} 
shows the temporal evolution of the flux-tube cross sections 
at $z=10~{\rm Mm}$ and $z=20~{\rm Mm}$. 
After { about three driving periods}, 
small scale vortices start to develop near the tube boundary, 
indicating the onset of the KHI. 
Similar to previous models 
\citep[e.g.,][]{2017A&A...604A.130K,2019ApJ...870...55G,2019ApJ...883...20G,2023ApJ...949L...1G}, 
these vortices develop approximately symmetrically with respect to $y=0$.
In addition, the cross section at $z=20~{\rm Mm}$, 
shown in Figure~\ref{fig:crs}(b), 
appears to evolve into a more { turbulent-like} state than that at $z=10~{\rm Mm}$. 
This is likely related to the larger wave amplitude at higher altitudes, 
where the decrease in density caused by gravitational stratification amplifies the transverse motions. 
The enhanced boundary shear then facilitates the further development of KH vortices.

 

We then examine the properties of the propagating waves. 
Figure~\ref{fig:yz_slice}(a) shows the distribution of $v_x$ on the $x=0$ plane 
at different times. 
A characteristic V-shaped pattern can be identified at each given time. 
This pattern is a typical signature of mode coupling between the propagating 
kink wave and the resonantly generated Alfv\'enic motions near the tube boundary. 
A similar behavior has been reported in previous numerical studies by
e.g., \cite{2010ApJ...711..990P}.
Figure~\ref{fig:yz_slice}(b) shows the $z$-component of the Poynting flux,
\begin{equation}
S_z
=
\frac{1}{\mu_0}
\left[
v_z\left(B_x^2+B_y^2\right)
-
B_z\left(v_xB_x+v_yB_y\right)
\right].
\label{eq:poyntingflux}
\end{equation}
The Poynting flux distribution shows a similar V-shaped structure to that 
seen in $v_x$, 
indicating that the upward wave energy is also redistributed toward the 
tube boundary where mode coupling takes place. 
In most of the physical domain, 
$S_z$ remains positive, 
confirming that the dominant energy transport is upward,
as the waves propagate upwards.
It also suggests that the enhanced viscosity buffer effectively 
suppresses strong reflections from the closed upper boundary, 
since no clear downward propagating energy flux pattern is seen in the physical domain below $z=50$ Mm.

We note, 
however, 
that the absence of a pronounced negative $S_z$ 
does not rule out 
weak local counter-propagating Alfv\'enic components due to the gravitational stratification. 
The total Poynting flux is indeed dominated by the upward propagating kink wave and the forward Alfv\'enic component. 
Therefore, 
local partial standing behavior may appear mainly as a reduction or local modulation of $S_z$, 
rather than as a strong net downward energy flux. 
This motivates 
more detailed Els\"asser variables and phase relation analyses below.

\begin{figure*}[ht!]
	\centering
	\gridline{\fig{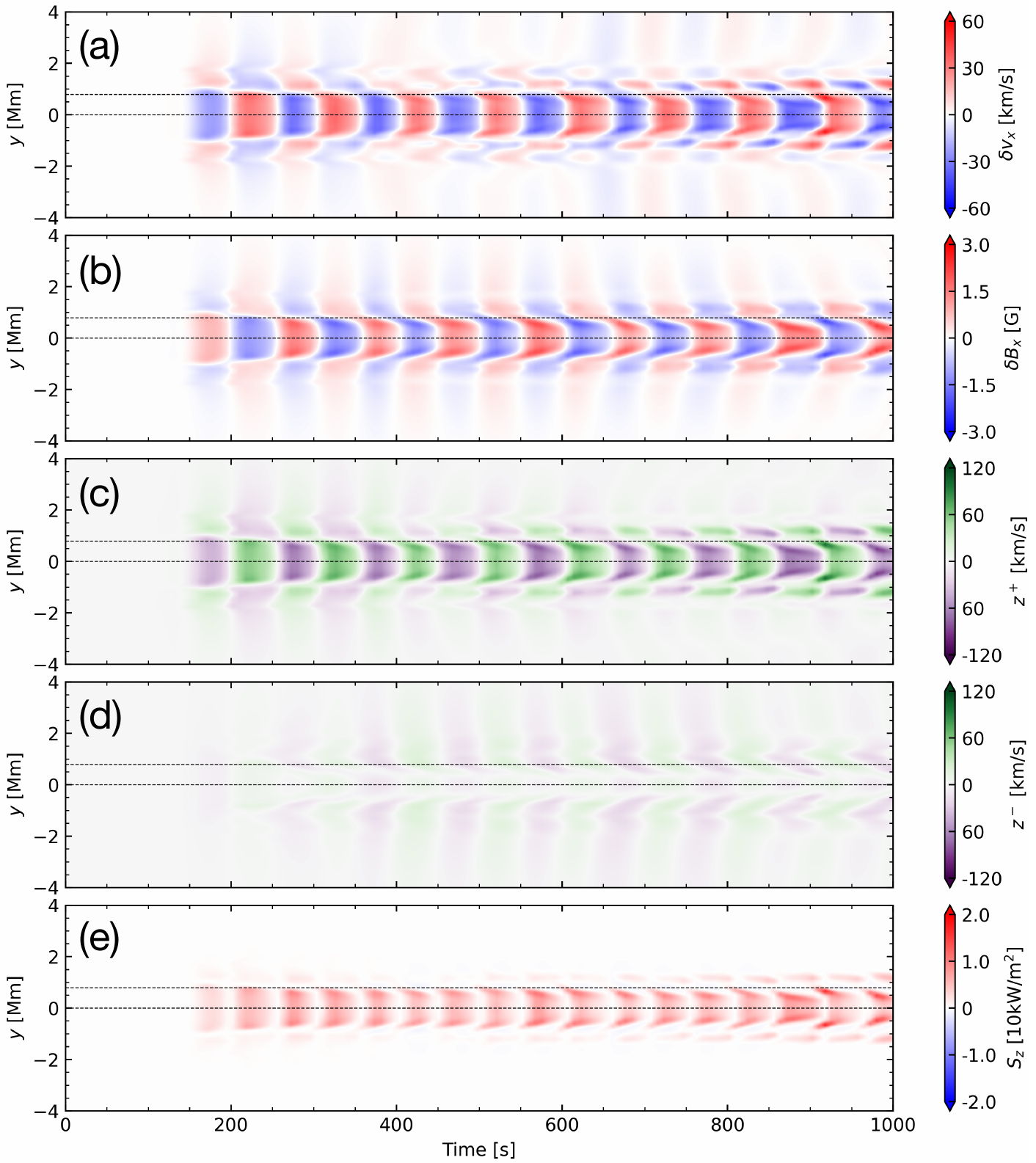}{0.7\textwidth}{}}
\caption{Temporal evolution of $\delta v_x$, $\delta B_x$, $z^{+}$, $z^{-}$, and $S_z$ along the $y$-axis 
at $x=0, z=20$ Mm. The dashed lines here indicate the positions at $y=0.8~{\rm Mm}$ and $y=0$, 
which are examined in more detail in Figure~\ref{fig:standing}.
\label{fig:tdmap}}
\end{figure*}

Figure~\ref{fig:tdmap}(a,b) shows the temporal evolution of the local velocity perturbation $\delta v_x$
and magnetic field perturbation $\delta B_x$ on the $x=0$ plane at the given height $z=20$ Mm. 
The wave signal reaches this height at about $t=150$ s, 
which is also consistent with the propagation seen in the animation of  {Figure~\ref{fig:crs}}. 
Inside the tube, 
$\delta v_x$ and $\delta B_x$ are mainly in anti-phase, 
which is the expected phase relation for an upward-propagating wave along the magnetic field line.
This anti-phase relation is also clearly seen at the tube axis, $y=0$, 
in Figure~\ref{fig:standing}(a1), 
and it persists throughout the entire computational time.
From about $t=300$ s onward, 
a V-shaped pattern develops in both $\delta v_x$ and $\delta B_x$ 
near the tube boundary. 
As discussed above, this feature is associated with resonant mode coupling
and therefore marks the development of boundary Alfv\'enic motions, 
while the tube-axis motion remains dominated by the propagating kink component.

The phase behavior becomes different near the tube boundary. 
In contrast to the tube axis, 
the boundary layer shows a more complex phase pattern after the onset of mode coupling. 
As shown in Figure~\ref{fig:standing}(b1), 
the phase difference between $\delta v_x$ and $\delta B_x$ becomes 
$\pi/2$ after about $t=300~{\rm s}$, 
which is close to the expected quadrature relation for standing Alfv\'enic motions \citep[see also,][]{2020ApJ...904..116G}.
This suggests that the resonantly generated Alfv\'enic motions near the boundary no longer 
behave as purely upward-propagating waves, 
motivating the following analysis of local standing behavior.

We then examine the possible standing waves near the tube boundary. 
Figure~\ref{fig:tdmap}(c)(d) shows the Els\"asser variables
$z^{\pm}$
on the $x=0$ plane at the fixed height of $z=20$ Mm. 
We employ $z^\pm$ as an ad hoc proxy for separating upwardly and downwardly propagating motions, 
where by ad hoc we acknowledge that the physical meaning of $z^\pm$ is not that clear
in compressible MHD \citep{1987JGR....92.7363M,2014ApJ...782...81V,2020ApJ...899..100V}. 
The upward propagating component $z^{+}$ is dominant 
throughout the tube cross section. 
Nevertheless, 
a weaker counter propagating component, 
$z^{-}$, can still be identified near the tube boundary, 
as shown in Figure~\ref{fig:tdmap}(d). 
This behavior is illustrated more clearly in Figure~\ref{fig:standing}(a2)(b2). 
Inside the tube, 
the $z^{+}$ component is much larger than $z^{-}$, 
consistent with the predominance of the propagating motions. 
Near the tube boundary, 
however, 
the two Els\"asser components become comparable in amplitude,
as shown in  Figure~\ref{fig:standing}(b2).
This indicates that the resonantly generated boundary Alfv\'enic motions no longer behave as purely upward propagating waves, 
but instead develop a local standing, or at least partial-standing property. 
This interpretation is further supported by the field aligned energy flux. 
As shown in Figure~\ref{fig:standing}(a3)(b3), 
the upward Poynting flux becomes much smaller near the tube boundary than at the tube axis. 
Such a reduction in the net upward energy flux is consistent with the presence of local counter propagating Alfv\'enic components 
and with the formation of local standing patterns in the boundary layer.

\section{Discussion} \label{sec:discussion}

\subsection{Local standing azimuthal Alfv\'en waves}\label{sec:discussion1_alf_standing}

One may ask why the resonantly transferred Alfv\'en waves near the tube boundary 
develop a locally standing character? 
To address this question, 
we revisit an important property of the present model, 
namely the atmosphere is gravitationally stratified, 
which produces a longitudinal variation of the local Alfv\'en speed. 
For the boundary Alfv\'enic motions, 
we define the local Alfv\'en wavelength on $x=0$ surface as
\begin{equation}
\lambda_{\rm A}(y,z)=v_{\rm A}(y,z)P ,
\label{eq:def_lambdaA}
\end{equation}
where \(P\) is the driving period. 
The characteristic length scale of the Alfv\'en speed variation is defined as
\begin{equation}
L_{\rm A}(y,z)=
\left|\frac{d\ln v_{\rm A}}{dz}\right|^{-1}
=
\left|\frac{v_{\rm A}}{d v_{\rm A}/dz}\right| .
\label{eq:def_LA}
\end{equation}
Owing to gravitational stratification, 
the Alfv\'en speed varies along the flux tube, 
producing a vertical Alfv\'en speed gradient.
This makes it possible for the local Alfv\'en wavelength, $\lambda_{\rm A}$, 
to become comparable to, or even larger than, the Alfv\'en speed variation scale, $L_{\rm A}$. 
When this occurs, 
the WKB approximation is no longer well satisfied, 
and non-WKB reflection can become non-negligible
{\citep[e.g.,][]{1980JGR....85.1311H,2005ApJS..156..265C}.}

\begin{figure}
	\centering
	\gridline{\fig{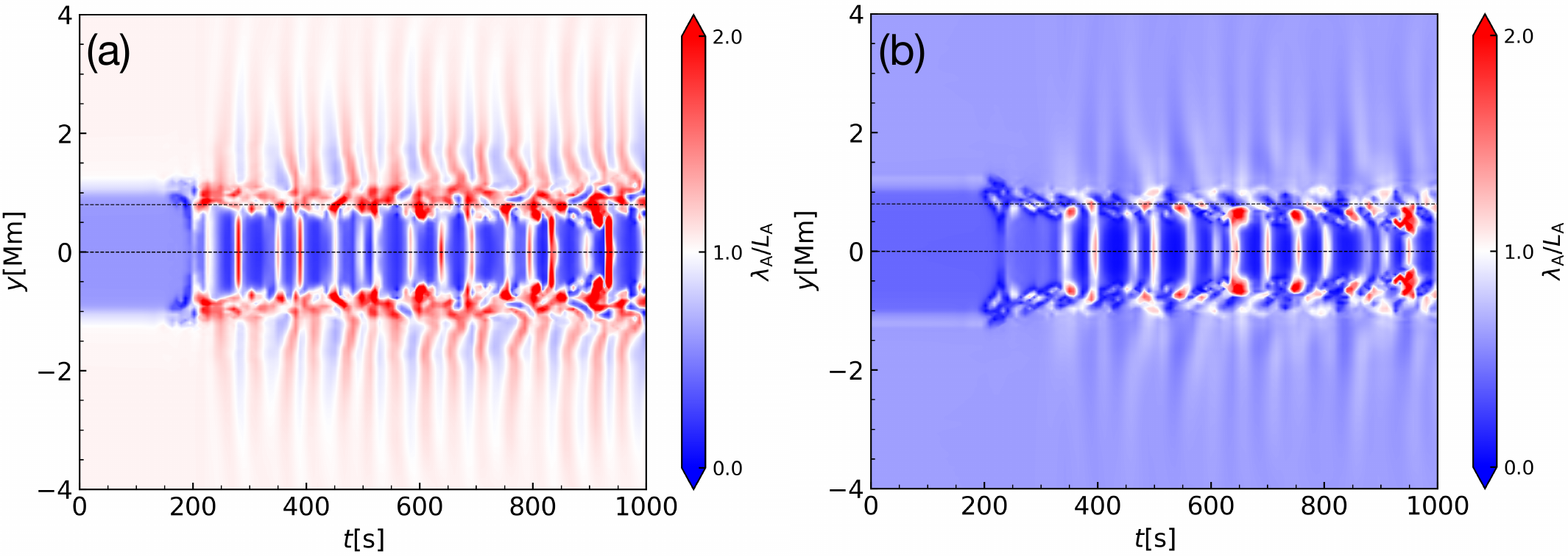}{1.0\textwidth}{}}
	\caption{
    Time distance maps of the ratio between the local Alfv\'en wavelength,
$\lambda_{\rm A}$, and the characteristic length scale of the Alfv\'en speed
variation, $L_{\rm A}$, along the $y$-direction at (a) $x=0$, $z=20~{\rm Mm}$ and
(b) $x=0$, $z=40~{\rm Mm}$. 
The black dashed lines indicate the positions at 
$y=0$ and $y=0.8~{\rm Mm}$.
}
	\label{fig:scale}
\end{figure}

Figure~\ref{fig:scale} shows the time distance distribution of $\lambda_{\rm A}/L_{\rm A}$. 
We compare this ratio at the tube axis, $y=0$, and near the tube boundary, $y=0.8~{\rm Mm}$. 
The region where $\lambda_{\rm A}/L_{\rm A}$ is larger than unity is more extended near the tube boundary than at the tube axis. 
This suggests that the non-WKB reflection is therefore more likely to occur near the tube boundary. 
This is consistent with the stronger counter-propagating Els\"asser component 
and the reduced field-aligned Poynting flux found there.
The ratio also shows a clear height dependence. 
At the higher height, $z=40~{\rm Mm}$, $\lambda_{\rm A}/L_{\rm A}$ becomes smaller than at lower heights, even near the tube boundary. 
This indicates that the WKB condition is better satisfied there, 
and hence the locally standing character of the boundary Alfv\'enic motions is expected to be weaker. 
To further examine this height dependence, 
we plot the temporal evolution of 
$\delta v_x$, $\delta B_x$, $z^{\pm}$, and $S_z$ at $z=40~{\rm Mm}$ in Figure~\ref{fig:standing_z40}. 
The phase difference between $\delta v_x$ and $\delta B_x$ deviates further from the quadrature relation expected for standing Alfv\'enic motions, 
and the amplitudes of $z^{+}$ and $z^{-}$ are no longer as comparable as those found at lower heights in Figure~\ref{fig:standing}.
Combining these diagnostics, 
we conclude that the resonantly generated boundary 
Alfv\'enic motions acquire a local standing character mainly in the regions where $\lambda_{\rm A}/L_{\rm A}$ is sufficiently large. 
These locally standing Alfv\'enic motions provide a more persistent velocity shear near the tube boundary 
and 
therefore play an essential role in promoting the development of the KHI.

\subsection{Local standing waves as a trigger for the KHI}\label{sec:discussion2_standing&KHI}

Our numerical results show that the KHI can develop near the tube boundary even 
in a flux tube whose global wave dynamics are dominated by propagating waves. 
The collective kink motion along the tube axis remains predominantly upward propagating, 
but the boundary dynamics are different. 
Near the tube boundary, 
transverse inhomogeneity leads to resonant mode coupling, 
through which energy from the propagating kink wave is transferred to localized Alfv\'enic motions. 
At the same time, 
longitudinal variation of the Alfv\'en speed allows weak non-WKB reflection to generate local counter-propagating Alfv\'enic components. 
As a result, 
the resonantly generated boundary Alfv\'enic motions acquire a locally standing character. 
As discussed by \citet{1983A&A...117..220H}, 
standing Alfv\'enic waves provide more favorable conditions for the development of the KHI. 
For purely propagating waves, 
the associated velocity shear is transported upward together with the wave packet. 
Once local counter-propagating Alfv\'enic components are present, 
however, 
the boundary Alfv\'enic motions can persist for a longer time at a given height, 
producing a more sustained shear layer. 
These local standing patterns therefore enhance the persistence of the boundary velocity shear and 
promote the development of KH vortices. 
Thus, 
propagating waves in an inhomogeneous open flux tube can still drive the KHI efficiently, 
provided that localized Alfv\'enic motions converted from kink waves attain a local standing state.

It is necessary to compare our results with previous propagating kink models. 
In \citet{2024A&A...689A.195G}, 
gravitational stratification and resonant absorption 
also lead to phase mixing and small-scale structures near the tube boundary, 
but the KHI remains limited and does not develop into a well developed turbulent state.
In fact, 
this difference does not necessarily contradict the non-WKB reflection argument. 
The condition that the local Alfv\'en wavelength becomes comparable to, 
or larger than, 
the Alfv\'en speed scale height only implies that counter propagating Alfv\'enic components may be generated. 
It does not by itself guarantee the formation of a coherent local standing wave. 
To form a local standing Alfv\'enic wave pattern, 
the counter propagating component has to become sufficiently strong and phase coherent with the forward component,
so that the net field-aligned energy flux is substantially reduced.
The different nonlinear evolution in the two models may be influenced by many factors. 
The model of \citet{2024A&A...689A.195G} considers a coronal open flux tube,
whereas the present model extends from the chromosphere to the corona. 
The amplitude of the velocity in the corona is thus much larger in the current model,
which can enhance the phase mixing and thus make non-WKB effects more pronounced. 
Aligning with this,
a larger amplitude of the velocity driver leads to the development of the phase mixed fine structures in the context of wave heating \citep[e.g.,][]{2017A&A...601A.107P}.
In addition,
it is also necessary to discuss the present results in the context of uniturbulence
\citep{2025A&A...696A.166V}.
In this scenario, 
transverse inhomogeneity itself can allow propagating waves to generate small-scale motions without requiring a globally standing wave pattern or strong counter-propagating wave packets. 
This idea is closely related to the current model, 
where the flux tube boundary provides the transverse inhomogeneity required for the generation of localized Alfv\'enic motions.
However,
uniturbulence emphasizes self-cascade caused by transverse structuring, 
our current results highlight the tranfer of propagating waves to local standing waves that are especially favorable for the KHI growth.

\subsection{Implications for dynamics in open filed regions of the corona }\label{sec:discussion3_implication}

The present scenario may be applicable to open magnetic structures in the solar atmosphere, 
such as chromospheric spicules and coronal hole plumes. 
As mentioned in the Introduction, 
observations suggest that such structures commonly host propagating kink like or Alfv\'enic motions, 
while their transverse structuring and gravitational stratification provide the physical conditions 
required for the mechanism explored in the present study. 

Direct detections of turbulent fine structures generated by propagating waves, 
however, remain challenging. 
In particular, 
small-scale KH vortices and boundary-layer turbulence may be difficult to resolve directly in current coronal observations.
Our results suggest that propagating Alfv\'enic waves may nevertheless leave 
observable spectroscopic signatures through unresolved nonlinear motions. 
Spectroscopic observations of open-field regions, especially polar coronal holes, 
have frequently reported enhanced non-thermal line widths, 
which are often interpreted in terms of unresolved Alfv\'enic motions
\citep[e.g.,][]{2013ApJ...776...78H, 2015NatCo...6.7813M}.
In the present model, 
propagating waves can generate locally standing Alfv\'enic motions near flux tube boundaries. 
These motions maintain persistent boundary shear and promote KH vortices, 
thereby producing unresolved small-scale velocities. 
Such nonlinear boundary motions may provide an additional source of 
unresolved velocities and therefore contribute to the observed nonthermal 
line broadening in coronal holes and plumes, 
beyond the contribution from the direct line-of-sight superposition of propagating Alfv\'enic waves.

It is also useful to relate the present result to the mixed-driver model of 
\citet{2019ApJ...870...55G} and to recent observations of 
torsional Alfv\'en waves. 
In \citet{2019ApJ...870...55G}, 
a mixed driver containing both transverse kink-like 
and torsional Alfv\'enic motions was shown to enhance the complexity 
of the boundary dynamics compared with a purely transverse driver. 
If a similar mixed driver were imposed in the present stratified 
open flux tube,
the additional torsional component would directly increase the azimuthal velocity 
shear near the tube boundary. 
Together with resonant absorption of the propagating kink wave, 
this would likely generate stronger localized Alfv\'enic motions, 
more efficient phase mixing, 
and more developed KH vortices. 
As a result, 
the system would be expected to evolve toward a stronger 
turbulent state than in the case driven by propagating kink waves alone.
This possibility is particularly relevant in light of the recent DKIST/Cryo-NIRSP 
observations reported by \citet{2026NatAs..10...42M}, 
which provided direct evidence for small-scale torsional Alfv\'en waves 
in the corona. 
These observations suggest that torsional Alfv\'enic motions may be a common component of coronal wave dynamics rather than 
a purely theoretical ingredient. 
Therefore, 
a realistic open magnetic structure may host both propagating kink-like 
motions and torsional Alfv\'enic perturbations. 
In such a situation, 
the mechanism identified here could become even more efficient,
namely propagating kink waves would supply energy to the boundary through resonant absorption, 
while torsional Alfv\'en waves would add direct azimuthal velocity shear. 
Their combined motion is expected to make the boundary layer more unstable, 
thereby producing more KH vortices and a more developed turbulent state.

\section{Summary} \label{sec:sum}

We perform a 3D MHD simulation of an open, 
gravitationally stratified magnetic flux tube extending from 
the chromosphere to the corona. 
Propagating kink waves are excited at the bottom boundary. 
Although propagating transverse waves are generally considered 
less favorable for the development of the KHI, 
we find that well developed KH vortices can still form in the current model. 
This is because resonant absorption transfers kink wave energy to 
azimuthal Alfv\'enic motions near the tube boundary,
while longitudinal stratification enables weak non-WKB reflection 
and allows these boundary motions to acquire a locally standing character. 
Our results suggest that propagating Alfv\'enic waves in open magnetic structures can drive turbulent evolution once local standing Alfv\'enic patterns are established near the boundary of the magnetic structure.

{
A further limitation of the present model is the absence of flux tube expansion. Realistic open structures generally expand with height, 
resulting in variations in the magnetic field strength, density distribution, and cross-sectional area along the tube. 
Such expansion would modify the longitudinal Alfv\'en speed profile and hence the spatial distribution of the Alfv\'en speed gradient that controls non-WKB reflection \citep{2005ApJS..156..265C}. 
Depending on the relative variations of the magnetic field and density, 
expansion may either enhance or reduce the local Alfv\'en speed gradient and shift the heights at which counter-propagating Alfv\'enic components are generated. 
It may also modify the wave amplitude, 
transverse structuring, and efficiency of resonant mode coupling near the tube boundary \citep[e.g.,][]{2019A&A...631A.105H}. 
Consequently, the heights at which locally standing Alfv\'enic motions develop,
as well as the onset and growth of the resulting KHI, may differ quantitatively in an expanding flux tube. 
The straight, uniform field configuration adopted here was chosen to isolate the effects of gravitational stratification and transverse density structuring. Nevertheless, the qualitative mechanism identified in this study remains applicable whenever the resulting longitudinal variation of the Alfv\'en speed is sufficiently strong to generate local counter-propagating Alfv\'enic components. 
Extending the current model to an expanding open flux tube can be a potential subject in the future.
}

\begin{acknowledgments}
We thank the referee for constructive comments that improved the manuscript.
This work is supported by the National Natural Science Foundation of China (12203030, 12373055). M.G. also acknowledges the support from the QILU Young Scholars Program of Shandong University, the Taishan Scholars Program Special Fund (tsqn202408051) and the Shandong Provincial Natural Science Foundation for Excellent Young Scientists Program, Overseas (2025HWYQ-019).
\end{acknowledgments}

\begin{figure*}[ht!]
	\centering
	\gridline{\fig{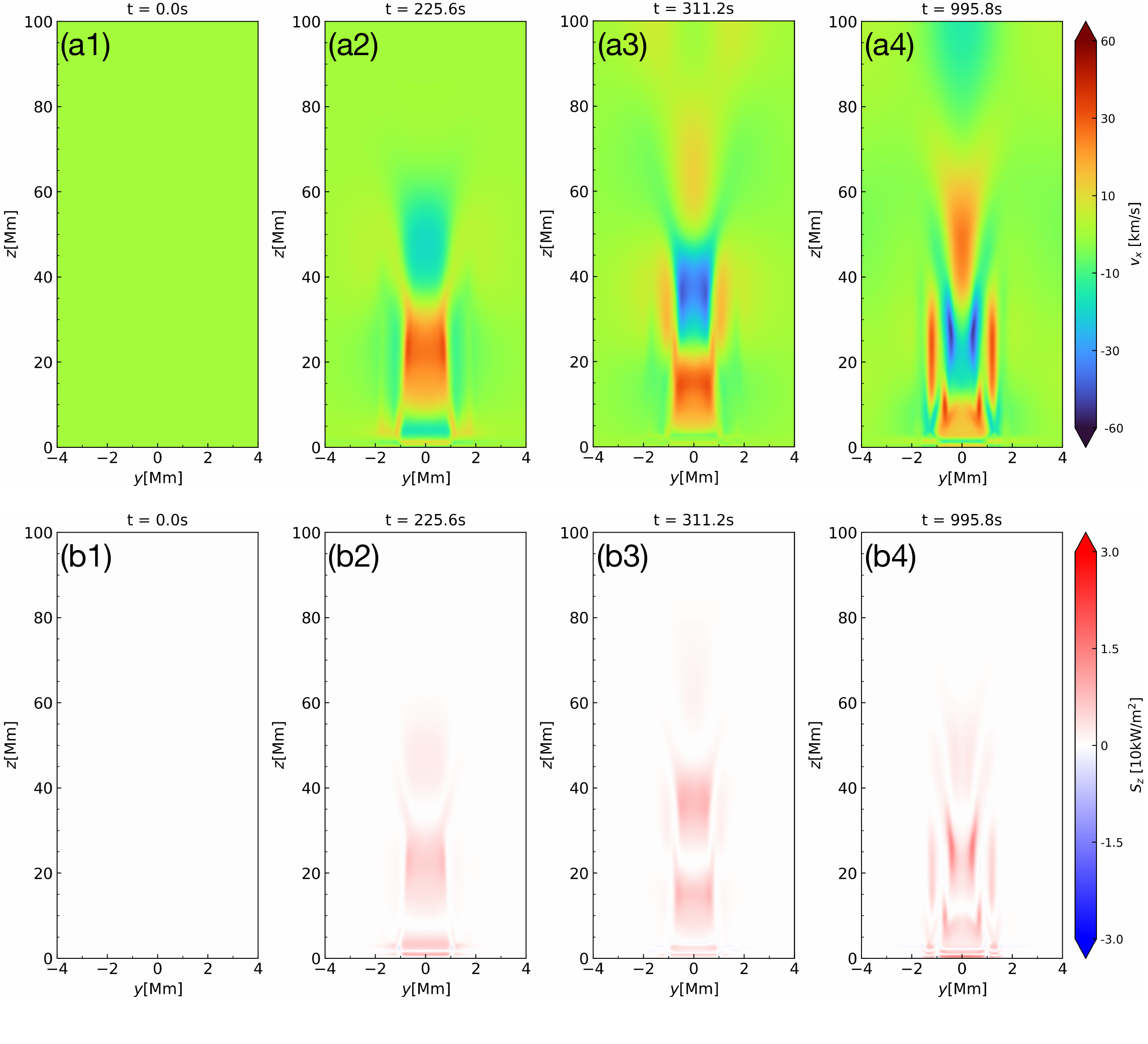}{0.8\textwidth}{}}
\caption{Snapshots of (a) $v_x$ and (b) the $z$-component of the Poynting flux, $S_z$, on the $x=0$ plane at the selected times as labelled.
\label{fig:yz_slice}}
\end{figure*}

\begin{figure*}[ht!]
	\centering
	\gridline{\fig{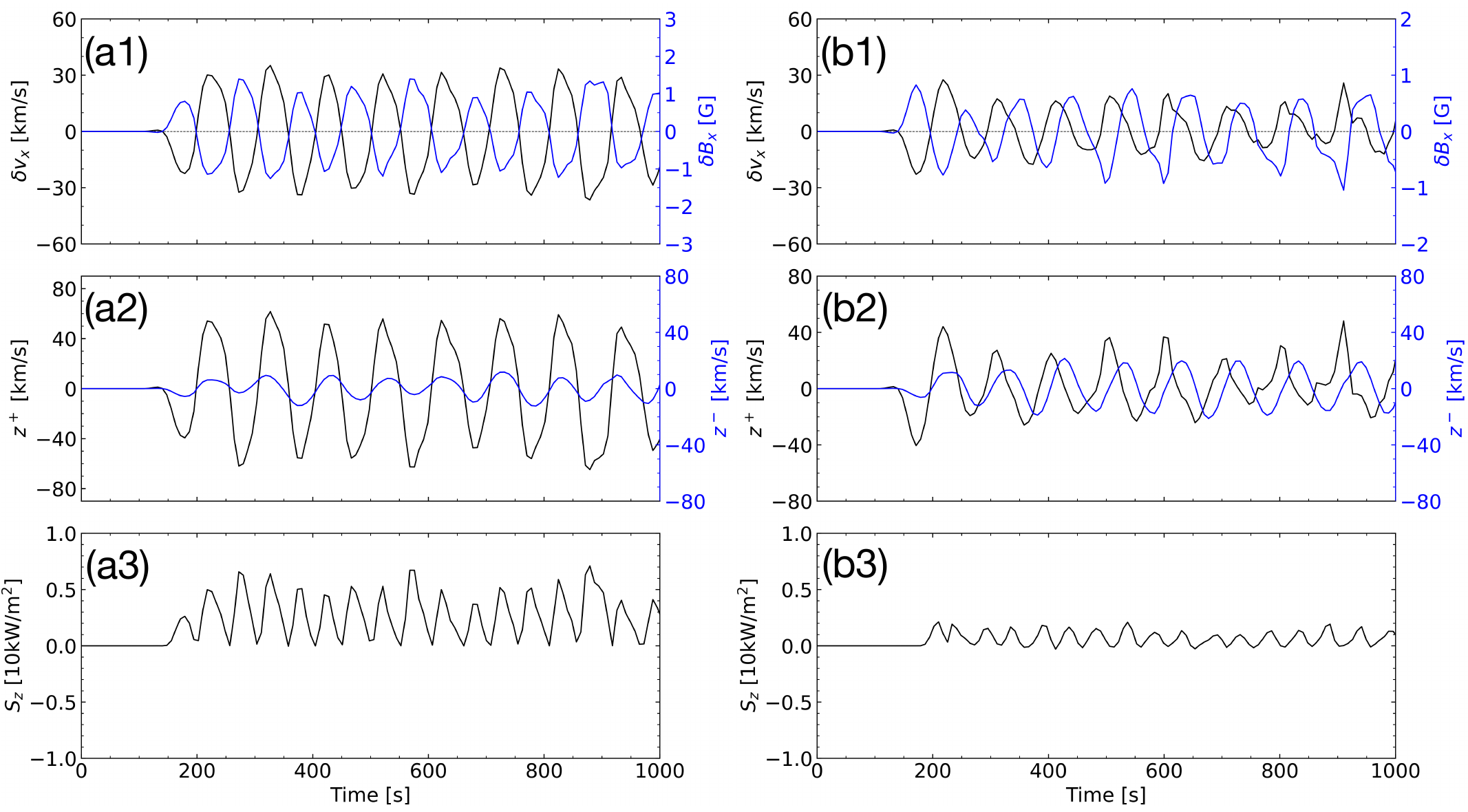}{1.0\textwidth}{}}
\caption{Temporal evolution of $\delta v_x$, $\delta B_x$, $z^{+}$, $z^{-}$, and $S_z$ at the two positions marked by the dashed lines in Figure~\ref{fig:tdmap}. 
Left panels: results at the tube axis, $y=0$. 
Right panels: results near the tube boundary, $y=0.8~{\rm Mm}$.
\label{fig:standing}}
\end{figure*}

\begin{figure}
	\centering
	\gridline{\fig{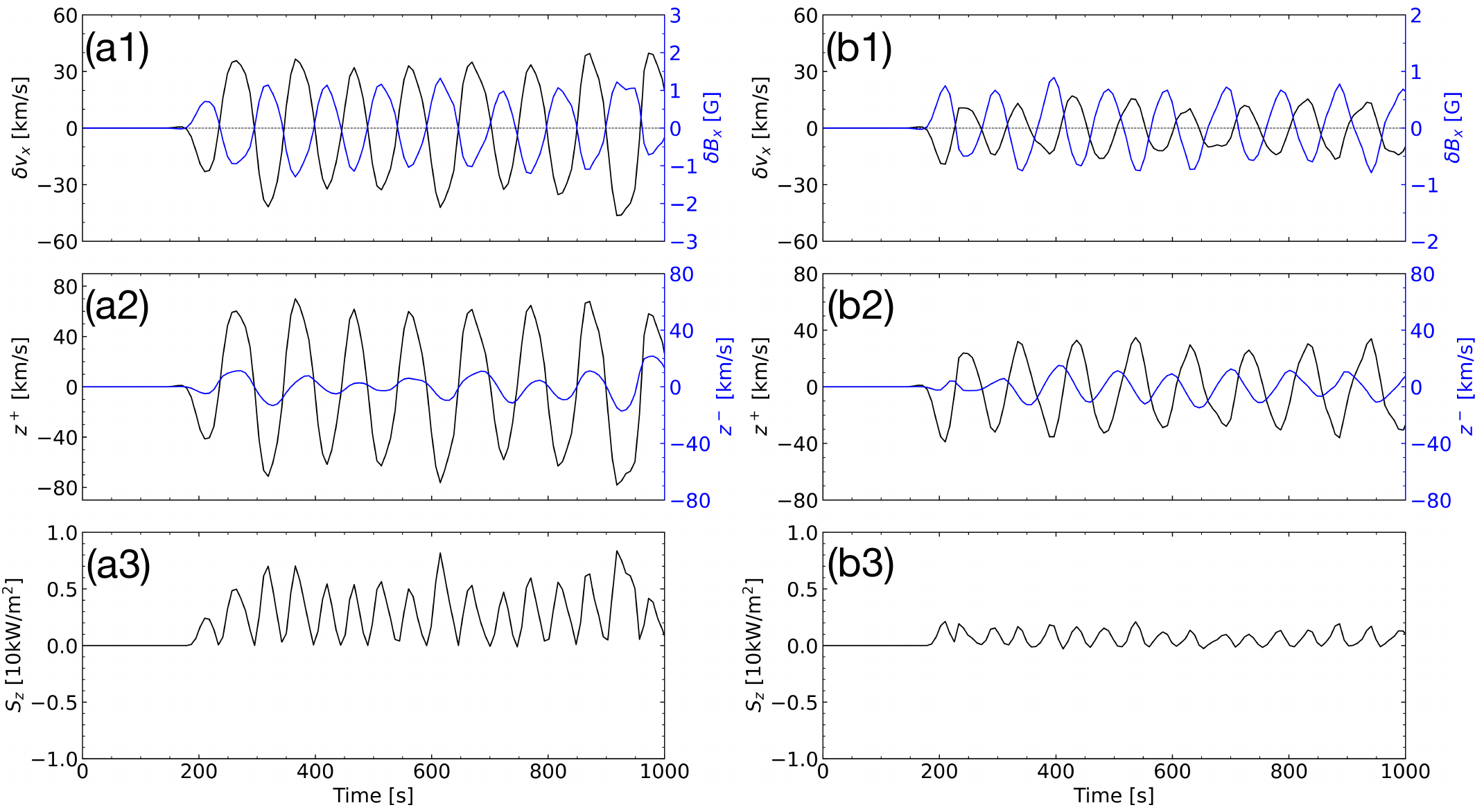}{1.0\textwidth}{}}
	\caption{
	Similar to Figure~\ref{fig:standing}, but for $z=40$ Mm.
	}
	\label{fig:standing_z40}
\end{figure}

\begin{figure}
	\centering
	\gridline{\fig{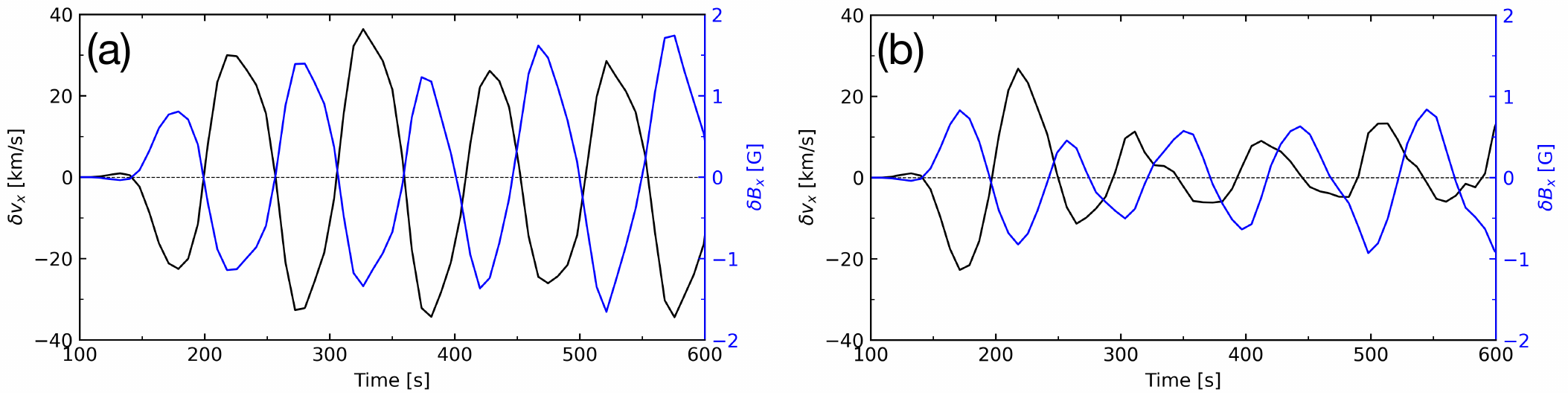}{1.0\textwidth}{}}
	\caption{
    {
 Temporal evolution of $\delta v_x$ and $\delta B_x$ at 
(a) $(x,y,z)=(0,0,20)~{\rm Mm}$ and 
(b) $(x,y,z)=(0,0.8,20)~{\rm Mm}$ 
in the test run with 
$z_{\rm buf}=60~{\rm Mm}$.
}
}
	\label{fig:standing2}
\end{figure}

\begin{figure}
	\centering
	\gridline{\fig{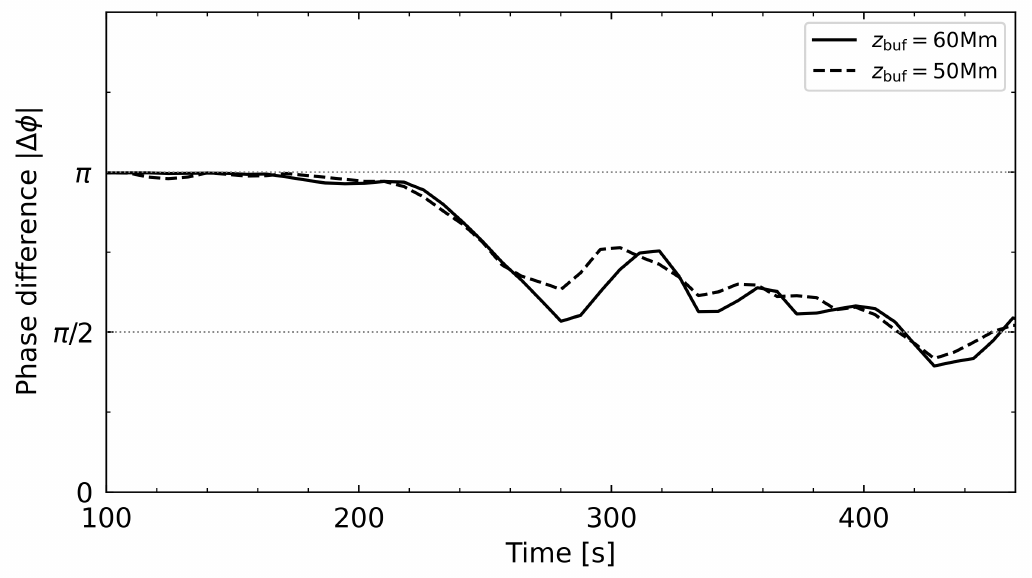}{0.5\textwidth}{}}
	\caption{
    {
   Temporal evolution of the absolute phase difference between $\delta v_x$ and 
$\delta B_x$ at $(x,y,z)=(0,0.8,20)~{\rm Mm}$ for the test run with 
$z_{\rm buf}=60~{\rm Mm}$ (solid line) and the reference run in the main text with 
$z_{\rm buf}=50~{\rm Mm}$ (dashed line).
}
}
	\label{fig:phase_diff}
\end{figure}

\clearpage
{
\appendix

\section{Assessment of the upper dissipative buffer and potential reflection}
\label{app}

The upper part of the computational domain contains an enhanced-viscosity layer designed to absorb upward-propagating motions before they reach the closed upper boundary. 
This layer should be regarded as an absorbing region rather than as a perfectly non-reflecting boundary. 
Because the interpretation of our results relies on the formation of locally standing Alfv\'enic motions near the tube boundary, 
we assess here whether possible residual reflection from the dissipative layer could account for the inferred counter-propagating components.

Preliminary tests with different dissipative buffer locations and viscosity values were performed to identify a configuration that maintains a stable atmosphere after relaxation. 
For the fixed tube length adopted in the main text, 
further reducing the Reynolds number did not produce a clear additional improvement in the absorption efficiency, 
but substantially increased the computational cost. 
Nevertheless, for comparison, we performed an additional simulation in which the start height of the dissipative buffer was shifted to $z_{\rm buf}=60~{\rm Mm}$ (see Equation~\eqref{eq_buffer}), 
while all other model parameters were kept unchanged.
As shown in Figure~\ref{fig:standing2}, 
the boundary Alfv\'enic motions again develop a locally standing character, 
whereas the collective motion at the tube axis remains propagating, 
consistent with the reference run shown in Figure~\ref{fig:standing}. 
A Hilbert transform analysis was further applied to the $\delta v_x$ and $\delta B_x$ signals shown in Figure~\ref{fig:standing2}(b) and Figure~\ref{fig:standing}(b1) in the main text, 
and the resulting phase difference is presented in Figure~\ref{fig:phase_diff}. 
The transition from the propagating wave phase relation, 
with an absolute phase difference close to $\pi$, 
to the locally standing wave relation, 
with an absolute phase difference close to $\pi/2$, 
occurs at nearly the same time in the two simulations. 
This comparison indicates that the onset of the locally standing boundary motions is insensitive to the buffer location.

Reflections would be expected to produce
downward-propagating signals affecting both the tube axis and boundary. 
However, 
the counter-propagating component found in the present simulation is enhanced 
and spatially localized near the inhomogeneous boundary layer, 
whereas the collective kink motion at the tube axis remains predominantly upward propagating. 
Moreover, the inferred standing signatures exhibit a height dependence,
namely stronger in regions where $\lambda_{\rm A}/L_{\rm A}$ is relatively large 
and become weaker at locations where the WKB condition is better satisfied. 
These spatial localization and height dependence are more consistent with distributed 
non-WKB reflection within the physical domain than with a reflection from the upper boundary.

Together with the predominantly upward Poynting flux shown in Figure~\ref{fig:yz_slice}, 
these results indicate that the enhanced viscosity buffer effectively suppresses strong reflection from the closed top boundary. 
Although a weak residual reflected component cannot be completely excluded, 
it is insufficient to account for the boundary-localized standing Alfv\'enic motions identified in the physical domain.
We therefore conclude that residual upper region reflection is unlikely to be the primary origin of the locally standing motions reported in the current study.
}

\clearpage

\clearpage

\clearpage

\clearpage

\clearpage

\clearpage

\clearpage

\clearpage

\clearpage
\bibliography{ref}{}
\bibliographystyle{aasjournalv7}



\end{document}